\documentclass[english]{emulateapj}
\usepackage[T1]{fontenc}
\usepackage{graphicx}
\usepackage{amssymb}

\makeatletter
\usepackage{apjfonts}
\slugcomment{}
\shorttitle{CAN THE LYC LEAKED OUT OF \ion{H}{2} REGIONS EXPLAIN THE DIG?}
\shortauthors{SEON}

\makeatother

\usepackage{babel}

\begin{document}

\title{Can the Lyman Continuum leaked out of \ion{H}{2} regions explain
Diffuse Ionized Gas?}

\author{Kwang-Il Seon\altaffilmark{1}}

\altaffiltext{1}{Korea Astronomy and Space Science Institute, Daejeon, Republic of Korea, 305-348; email: kiseon@kasi.re.kr} 
\begin{abstract}
We present an attempt to explain the diffuse H$\alpha$ emission of
a face-on galaxy M 51 with the ``standard'' photoionization model,
in which the Lyman continuum (Lyc) escaping from \ion{H}{2} regions
propagates large distances into the diffuse interstellar medium (ISM).
The diffuse H$\alpha$ emission of M 51 is analyzed using thin slab
models and exponential disk models in the context of the ``on-the-spot''
approximation. The scale height of the ionized gas needed to explain
the diffuse H$\alpha$ emission with the scenario is found to be of
the order of $\sim1-2$ kpc, consistent with those of our Galaxy and
edge-on galaxies. The model also provides a vertical profile, when
the galaxy is viewed edge-on, consisting of two-exponential components.
However, it is found that an incredibly low absorption coefficient
of $\kappa_{0}\approx0.4-0.8$ kpc$^{-1}$ at the galactic plane,
or, equivalently, an effective cross-section as low as $\sigma_{{\rm eff}}\sim10^{-5}$
of the photoionization cross-section at 912\AA\ is required to allow
the stellar Lyc photons to travel through the \ion{H}{1} disk. Such
a low absorption coefficient is out of accord with the properties
of the ISM. Furthermore, we found that even the model that has the
DIG phase only and no \ion{H}{1} gas phase shows highly concentrated
H$\alpha$ emissions around \ion{H}{2} regions, and can account for
only $\lesssim26$\% of the H$\alpha$ luminosity of the DIG. This
result places a strong constraint on the ionizing source of the DIG.
We also report that the H$\alpha$ intensity distribution functions
not only of the DIG, but also of \ion{H}{2} regions in M 51, appear
to be lognormal.
\end{abstract}

\keywords{galaxies: individual: M 51 --- galaxies: ISM --- ISM: \ion{H}{2}
regions --- ISM: structure}

\section{Introduction}

Diffuse ionized gas (DIG) is a major component of the interstellar
medium (ISM) in galaxies \citep{Reynolds91,Walterbos96,Haffner2009}.
The large energy requirement of the DIG strongly suggests that OB
stars are the only viable ionization source \citep{Reynolds84}. The
``standard'' photoionization model of the DIG assumes that the ionizing
photons that leaked out of \ion{H}{2} regions in the galactic disk
ionize the diffuse ISM \citep{Mathis86,Sokolowski1991,Domgorgen94}.
If the Lyman continuum (Lyc, ionizing continuum) photons escaping
from the traditional \ion{H}{2} regions were the dominant source
responsible for ionizing the DIG, then the H$\alpha$ intensity from
the DIG would be expected to correlate with that of discrete \ion{H}{2}
regions. \citet{Ferguson96} found a correlation between the DIG and
bright \ion{H}{2} regions over both small and large scales.

However, it is not clear how a major fraction of Lyc photons can escape
from the immediate surroundings of an OB association and how the escaped
photons can be transmitted across distances of the order of a kpc.
\citet{Norman89} suggested that superbubbles or ``chimneys'' around
OB associations could ionize the halo gas above the chimneys. \citet{Miller93},
\citet{Dove94}, and \citet{Dove00} quantitatively investigated the
possibility of penetration of the Lyc into the Galactic halo. These
senarios may explain the diffuse H$\alpha$ emission, in particular
vertical H$\alpha$ filaments or ``worms,'' seen in external edge-on
galaxies, such as NGC 891 \citep{Rand1990,Dettmar1990}. However,
the Lyc must travel not only large distances above the plane, but
also within a galactic plane \citep{Reynolds93}. This requirement
seems evident when the diffuse H$\alpha$ emission far from the classical
\ion{H}{2} regions in face-on galaxies are examined.

\citet{Zurita02} were the first to attempt modeling the global morphology
of the DIG of a face-on galaxy with the Lyc escaping from bright \ion{H}{2}
regions. They assumed a thin slab disk with a constant absorption
coefficient over a galaxy NGC 157 and compared the predicted morphologies
with the observed data by varying the absorption coefficient. In this
way, they found that the diffuse H$\alpha$ surface brightness distribution
is, surprisingly, well reproduced by assuming no significant absorption
of ionizing photons.

However, they neither compared the absolute H$\alpha$ flux of the
models with the observed data nor discussed the implications of the
absorption coefficient obtained to explain the observed morphology.
If OB stars are to provide the diffuse ionization, then the ISM must
be not only transparent to ionizing photons, but also sufficiently
dense to provide the observed H$\alpha$ emission \citep{Miller93}.
In the extreme case of no absorption, no H$\alpha$ radiation is emitted.
Therefore, it is necessary to compare the absolute H$\alpha$ flux
as well as the morphology of the H$\alpha$ image. It is also worthwhile
to consider a geometrically thick disk in the modeling and compare
the scale height obtained to account for the observed H$\alpha$ luminosity
with those of edge-on galaxies.

The transmission of the Lyc into the DIG has been attributed to the
inhomogeneity of the ISM, such as a fractal \ion{H}{1} distribution
produced by ISM turbulence \citep{Elmegreen1997,Wood2004}. However,
no attempt has been made to quantify the ISM density structure required
to allow the transmission of the Lyc. It is not clear whether such
inhomogeneity of the ISM in typical spiral galaxies is consisitent
with the requirement of the ``photoionization'' scenario. Meanwhile,
it is known that the probability distribution functions (PDFs) of
not only the local densities, but also the column densities of the
turbulent ISM, are lognormal \citep{Vazquez94,Ostriker01,Fischera04}.
With this property, it is possible to quantitatively access how turbulent
or porous the ISM should be for the stellar Lyc radiation to be responsible
for the diffuse H$\alpha$ emission with the ``standard'' photoionization
model or whether the absorption coefficient obtained in the modeling
is in accord with the current knowledge of the ISM.

In this paper, we calculate thin slab models similar to those of \citet{Zurita02}
and more elaborated models incorporationg an exponential disk for
a face-on galaxy M 51. We find that the ``standard'' scenario requires
absorption too unrealistically small to be believed, but the obtained
scale-height of the galactic disk is consistent with those of edge-on
galaxies. We also report that the PDFs of the H$\alpha$ intensities
of the DIG and \ion{H}{2} regions in the galaxy M 51 are lognormal.

\section{H$\alpha$ Data}

For our purposes, a face-on galaxy is the best choise since we can
easily identify \ion{H}{2} regions, i.e., the ionizing source that
is expected to be the dominant source. When an H$\alpha$ emission
map is ready, the first step for the analysis is then to separate
\ion{H}{2} regions from the diffuse H$\alpha$ emission. To do so,
we used an automated photometry procedure known as ``HIIphot,'' described
in \citet{Thilker00}, in which they applied the procedure to the
H$\alpha$ data of M 51. Therefore, as a matter of convenience, we
use the same H$\alpha$ data of M 51 as they did. Their publicly available
data, provided together with the ``HIIphot'' procedure, has the same
angular resolution, but a slightly smaller field size than that of
\citet{Thilker00}.

For the present analysis, we assumed a distance of 9.6 Mpc to M 51
\citep{Sandage74}. At this distance, a pixel in the image corresponds
to a linear resolution of $\Delta r=32$ pc. No correction was attempted
for extinction to compare with earlier studies. We detected 1356 \ion{H}{2}
regions. Of the total sample, 1061 regions were classified as ``photometric-quality''
detections having S/N $\ge$ 5. The extent of each \ion{H}{2} region
was determined using a terminal surface brightness slope of 1.5 EM
pc$^{-1}$, as in \citet{Thilker00}. Here, EM is an emission measure
(cm$^{-6}$ pc). A continuum-subtracted H$\alpha$ image, where \ion{H}{2}
regions were masked, and the H$\alpha$ intensity histograms for the
DIG and \ion{H}{2} regions are shown in Figure \ref{m51_map}. It
is evident in the figure that the H$\alpha$ intensity histograms
are lognormal, both in the DIG and \ion{H}{2} regions. The width
of the lognormal distribution determines the porosity of the medium
and is essential to the analysis of the transmission of the Lyc. The
fit results will be used for the modeling of the DIG and discussed
in later sections.

\section{Models}

\subsection{Basic Assumptions}

We employed the ``on-the-spot'' approximation, in which ionizing photons
emitted by recombinations to the ground state of hydrogen in the ionized
gas are assumed to be reabsorbed very close to their point of origination.
However, strictly speaking, the approximation may not be applicable
to the optically thin case that is the primary concern of the present
study. The probability of a photon being remitted as an ionizing photon
is the ratio of the recombination coefficient to the ground state
of hydrogen ($\alpha_{1}$) and that to all levels ($\alpha_{{\rm A}}$).
The ratio is $0.33-0.43$ for a gas in a temperature range of $5000-20,000$
K \citep{Osterbrock06}. This portion of the diffuse ionizing photons
may escape the system or be reabsorbed elsewhere in the system. Therefore,
the ``on-the-spot'' approximation would overestimate the total H$\alpha$
intensity by a factor of $\lesssim2$ in the optically thin case.
The H$\alpha$ morphology would also be probably twice sharper than
that of a more accurate calculation. These two limitations should
be kept in mind when interpreting our results.

The purpose of this study is to simulate the situation in which the
neutral hydrogen gas has a vertical scale height of $h$, and a density
structure that can be described by an absorption coefficient $\kappa(z)$
in the global sense so that some of the neutral hydrogen is ionized
by the Lyc radiation. Since we assume that the Lyc photons absorbed
in a volume are balanced by the immediate recombination, the H$\alpha$
surface brightness at each pixel is directly proportional to the absorbed
number of Lyc photons within the volume and is independent of the
unresolved cloud density structure, such as clumpiness or porosity.
The amounts of Lyc intensities, which are emitted from all \ion{H}{2}
regions and absorbed at a point in the DIG, are co-added to calculate
the H$\alpha$ intensity at the point.

\subsection{Thin Slab Models}

We first calculated two-dimensional thin slab models using the method
described in \citet{Zurita02}, not only in order to compare our results
with theirs, but also to obtain basic ideas on the parameter ranges
needed to calculate more elaborated models. The \ion{H}{1} layer
in the galactic disk is assumed to be a thin slab of a constant thickness
$\Delta H=2h$ and to have a constant absorption coefficient $\kappa$
(kpc$^{-1}$). We took an average \ion{H}{1} scale height of $h=100$
pc. We calculated the total radiation field absorbed within a volume
$\Delta V=(\Delta r)^{2}\Delta H$ corresponding to a pixel of the
H$\alpha$ image of the face-on galaxy. Here, $(\Delta r)^{2}$ is
the area corresponding to a pixel in the image, and the area of the
volume subtended by the source $\Delta A=\Delta H\Delta r$. Because
of geometrical dilution and absorption, the transmitted flux of stellar
radiation at distance $r$ from a source is $L_{0}e^{-\tau}/4\pi r^{2}$,
where $L_{0}$ is the total number of ionizing photons emitted per
second at the source, and the optical depth $\tau=\int\kappa dr$.
The stellar luminosity absorbed over volume $\Delta V$ after traveling
the optical depth $\tau$ is then given by $e^{-\tau}\Delta\tau L_{0}(\Delta A/4\pi r^{2})$,
or, in general, $e^{-\tau}(1-e^{-\Delta\tau})L_{0}(\Delta A/4\pi r^{2})$,
where $\Delta\tau=\kappa\Delta r$.

Various absorption coefficients are examined to account for the effect
of density inhomogeneity. We consider each pixel of \ion{H}{2} regions
as a point source with the effective Lyc luminosity equal to that
corresponding to its measured H$\alpha$ intensity. Given a constant
absorption coefficient $\kappa$, we obtain the H$\alpha$ surface
brightness at pixel $i$:

\[
I_{i}^{{\rm DIG}}=\sum_{j}\frac{f_{{\rm scale}}I_{j}^{{\rm HII}}}{4\pi r_{ij}^{2}}\exp\left(-\kappa r_{ij}\right)\Delta A(1-e^{-\kappa\Delta r}),\]
where $f_{{\rm scale}}$ is the ratio of the total luminosity of DIG
to that of \ion{H}{2} regions. A large fraction of the total H$\alpha$
emission (DIG + \ion{H}{2} regions), $\sim0.5$, is contributed by
DIG, implying that the total Lyc luminosity emitted from the sources
is about twice higher than that absorbed in \ion{H}{2} regions. The
luminosities from the DIG and \ion{H}{2} regions are then approximately
equal to each other ($f_{{\rm scale}}\approx1$). Here, $r_{ij}$
is the distance from a source pixel $j$ to the pixel $i$, and $\tau_{ij}=\kappa r_{ij}$
is the optical depth along the path from the pixel $i$ to $j$. The
conversion factors, the Lyc to H$\alpha$ luminosity and the luminosity
to flux, are canceled out on both sides of the equation.

It should be noted here that \citet{Zurita02} ignored the factor
$\Delta A(1-e^{-\kappa\Delta r})$. The effect of the factor $1-e^{-\kappa\Delta r}$
$(\approx\kappa\Delta r$ as $\kappa\Delta r\ll1)$ is to lower the
amount of produced H$\alpha$ emission as the medium becomes increasingly
transparent to the Lyc. The H$\alpha$ brightness increases with the
disk thickness $\Delta H$. The factor $f_{{\rm scale}}$ is allowed
to vary to take into account a larger scale height as well as to match
the observed data, although it is initially meant to be the ratio
of Lyc leakage toward DIG to that absorbed in \ion{H}{2} regions.
In other words, the scale height needed to be in accord with the observed
data is $h\approx0.1f_{{\rm scale}}$ kpc. In this way, the scale
height of the galaxy can be inferred.

We have varied the effective absorption coefficient from $\kappa=10^{-3}$
to $10^{2}$ kpc$^{-1}$. Typical results are shown in Figure \ref{model_samples}.
The scaling factors, defined by $f_{{\rm scale}}\equiv\sum I_{{\rm data}}/\sum I_{{\rm model}}$,
were multiplied to the models to match the overall H$\alpha$ brightness
level. We also show the intensity distributions of the model, observed
data in the second row, and correlation plots between the model and
observed flux in the form of two-dimensional histograms in the third
row.

We note that, at the optically thin limit of $\kappa\lesssim0.1$
kpc$^{-1}$, as the attenuation factor $e^{-\kappa r}\approx1$ for
a system size of $\sim10$ kpc, the morphologies of the models are
basically the same as in the case of no absorption $(\kappa=0)$ and
are determined by the geometrical dilution only. In the limit, however,
we obtain much lower H$\alpha$ intensity than the observed data although
the overall morphology seems, at first glance, to resemble the observed
image. The model and observed intensities are, in a general sense,
linearly proportional to each other, as shown for $\kappa\approx0.1$
kpc$^{-1}$ in the third row of Figure \ref{model_samples}. However,
the model images are a bit broader near \ion{H}{2} regions than the
observed image, as also noted by \citet{Zurita02}. As $\kappa$ increases,
the overall morphology begins to deviate from the observed one. For
large absorption coefficients (e.g., for $\kappa=3.16$ kpc$^{-1}$),
the H$\alpha$ emissions are mostly concentrated near \ion{H}{2}
regions. The correlation between the model and observed data also
begins to deviate from the linear proportionality, as can be noted
in third row of Figure \ref{model_samples}. We, therefore, fitted
the correlation between the observed and model intensities with a
power-law $I_{{\rm data}}=f_{{\rm scale}}^{*}\left(I_{{\rm model}}\right)^{\beta}$.

\begin{figure}[!t]
\begin{centering}
\includegraphics[clip,scale=0.4]{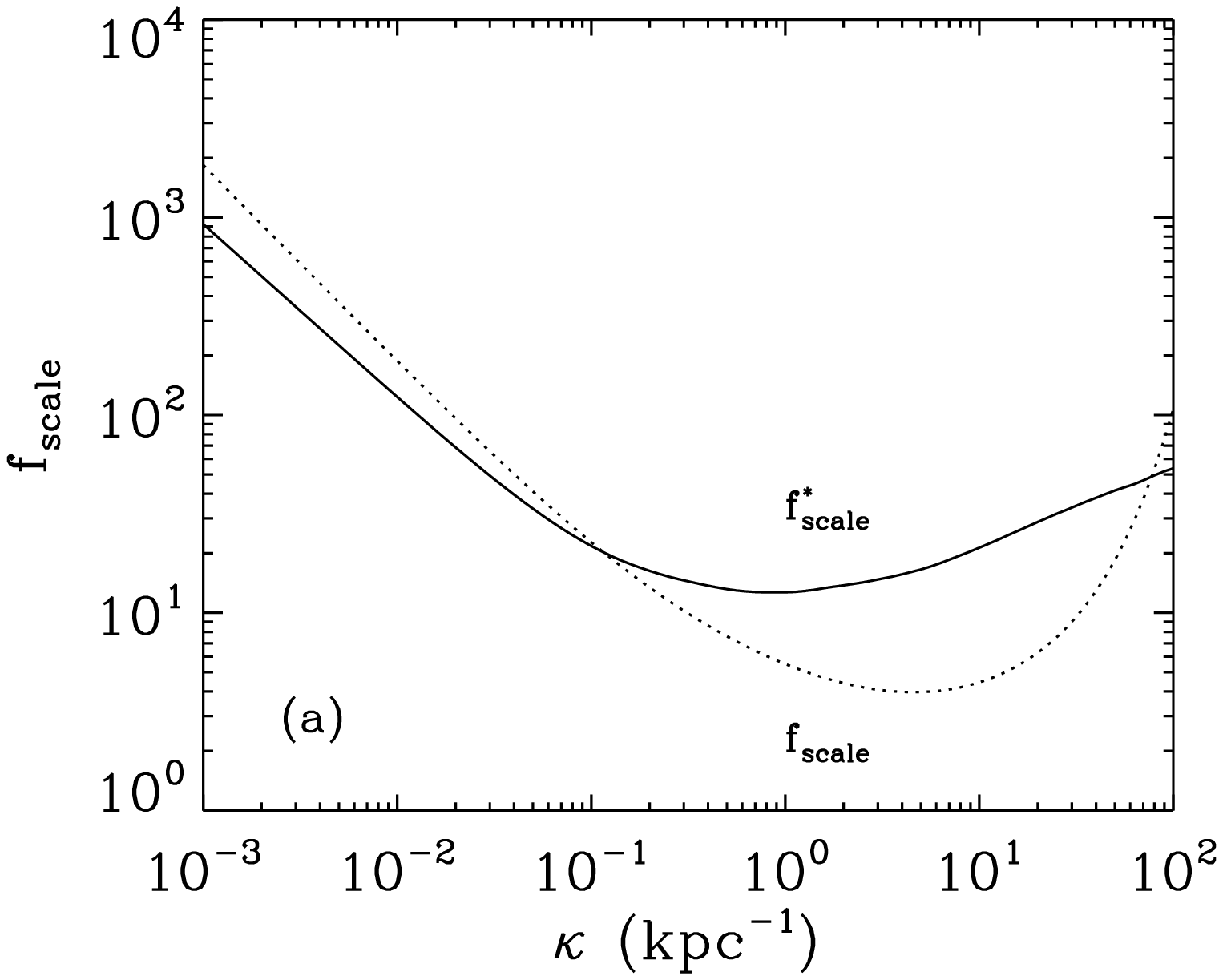}
\par\end{centering}

\begin{centering}
\includegraphics[clip,scale=0.4]{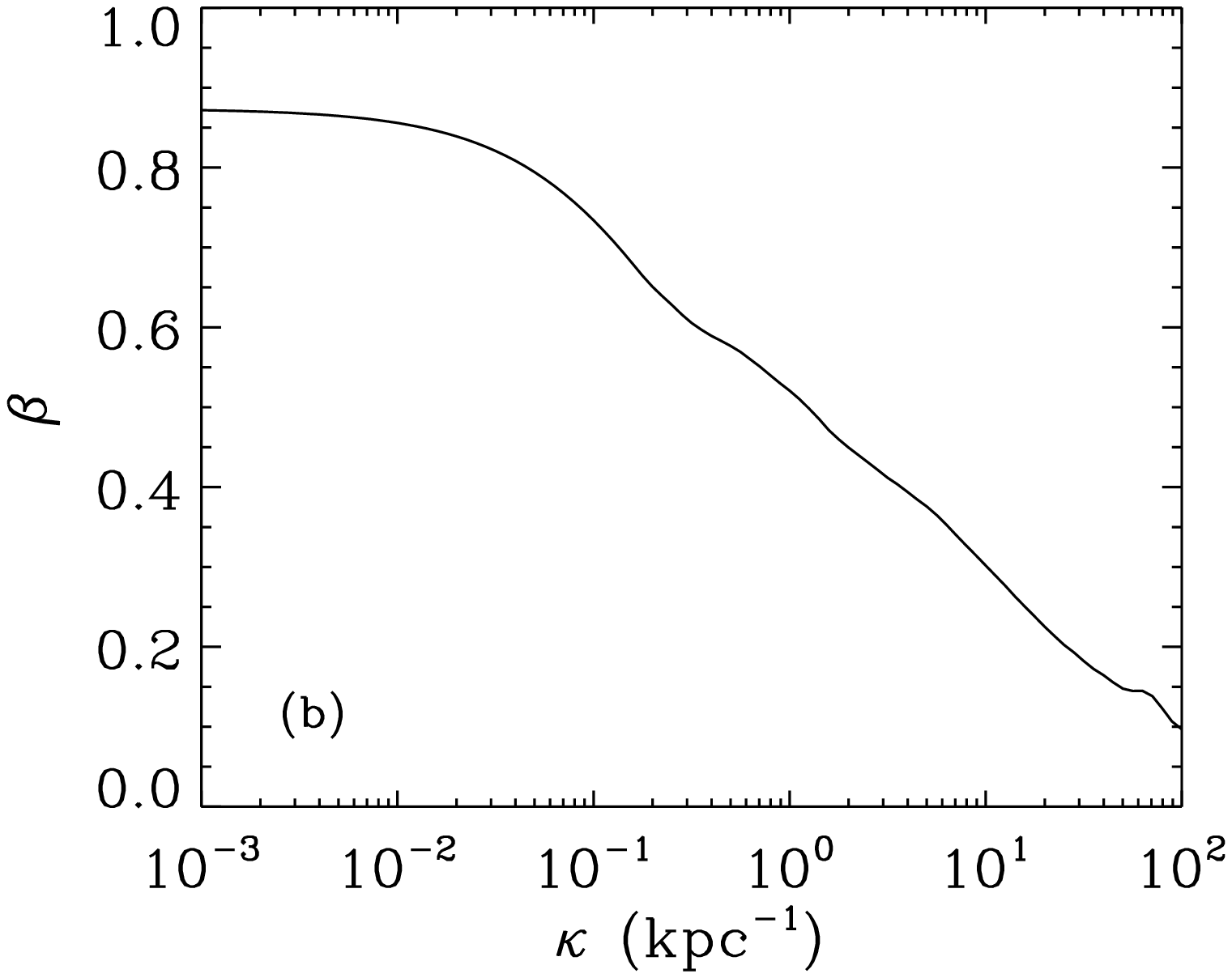}
\par\end{centering}

\caption{\label{model_fit}Best fit parameters with $I_{{\rm data}}=f_{{\rm scale}}^{*}\left(I_{{\rm model}}\right)^{\beta}$.
The dotted line was obtained by $f_{{\rm scale}}=\sum I_{{\rm data}}/\sum I_{{\rm model}}$.}

\end{figure}

The results are shown in Figure \ref{model_fit}, together with $f_{{\rm scale}}$.
The third row of Figure \ref{model_samples} shows the best fit curves
in red as well as the one-to-one curves in blue. In Figure \ref{model_fit}(b),
the slope $\beta$ is a little smaller than 1 even at $\kappa\ll0.1$
kpc$^{-1}$. It is also shown that the slope $\beta$ decreases rapidly
with increasing $\kappa$ as $\kappa\gtrsim0.1$, implying that the
observed morphology cannot be reproduced well in this range because
of fast attenuation of the Lyc while it is being propagated into the
diffuse ISM. However, when the correlation histograms are compared
with blue one-to-one curves, it is found that the correlation slope
is close to 1 up to $\kappa=0.2$ kpc$^{-2}$. Meanwhile, in the limit
of $\kappa\lesssim0.1$ kpc$^{-1}$, the scale factor increases with
decreasing $\kappa$, as shown in Figure \ref{model_fit}(a). Therefore,
the optimal solution is obtained with $\kappa\approx0.1-0.2$ kpc$^{-1}$.
We also note that large scaling factors of $\sim13-22$ are required
for these values, implying $h\sim1-2$ kpc for the generally accepted
Lyc leakage fraction of $\sim0.5$. The required scale height would
be $\sim1-4$ kpc, considering that the present results may be overestimated
by up to a factor of $2$. These values are too large to be reconciled
with the \ion{H}{1} disk scale height, but consistent with the DIG
scale heights of $\sim1-3$ kpc, measured from our Galaxy and external
edge-on galaxies \citep[e.g.,][]{Reynolds89,Rand1996,Collins2001}.

A serious problem encountered in the present results is an extremely
low absorption coefficient. Assuming the same \ion{H}{1} column density
of $N_{{\rm HI}}=8.6\times10^{20}$ cm$^{-2}$ through the galactic
disk of $h=100$ pc as \citet{Zurita02}, we obtain the effective
cross-section $\sigma_{{\rm eff}}=\kappa/(N_{{\rm HI}}/\Delta H)=2.3\times10^{-23}$($\kappa/$0.1
kpc$^{-1}$) cm$^{2}$. \citet{Zurita02} obtained the best approximation
to the morphology of the observed DIG surface brightness with $\sigma_{{\rm eff}}=5.3\times10^{-23}$
cm$^{2}$. This value corresponds to $\kappa=0.2$ kpc$^{-1}$, consistent
with our result of $\kappa\approx0.1-0.2$ kpc$^{-1}$. Assuming a
smaller column density of $N_{{\rm HI}}=10^{20}$ cm$^{-2}$, we obtain
$\sigma_{{\rm eff}}=2\times10^{-22}$($\kappa/$0.1 kpc$^{-1}$) cm$^{2}$.
These cross-sections imply virtually no absorption of the Lyc, i.e.,
as low as $\sim10^{-6}-10^{-5}$ of the photoionization cross-section
$6.4\times10^{-18}$ cm$^{2}$ at 912\AA\ \citep{Verner95}. We also
note that $\sigma_{{\rm eff}}$ is even lower than the dust extinction
cross-section of $2.3\times10^{-21}$ cm$^{2}$/H at 912 \AA\ \citep{Draine03}.

\subsection{Exponential Disk Models}

For exponential disk models, we employ a three-dimensional Monte Carlo
radiation transfer code \citep{Wood1999} that was originally developed
for the simulation of dust-scattering. A modified version has been
used in the radiative transfer models of dust-scattering in the Ophiuchus
region \citep{Lee2008}. The code was modified to simulate photoionization
in the context of the ``on-the-spot'' aproximation. Lyc photons are
emitted from point source \ion{H}{2} regions, as in thin slab models.
We tracked the propagation of the Lyc photons from sources as they
were absorbed and reemitted as H$\alpha$ photons. The absorbed photon
was 100\% reemitted isotropically as H$\alpha$ photon in the present
simulations, while in dust-scattering simulations the photon was reemitted
with a probability of $0<a<1$ toward a new direction determined by
a Henyey-Greenstein scattering phase function. In other words, our
simulations were performed by assuming the albedo $a=1$, asymmetry
factor $g=0$, and a single scattering. The reemitted H$\alpha$ traveled
toward the observer and was added into the H$\alpha$ image. Absorption
coefficients in each density grid were calculated assuming an exponential
function, $\kappa\left(z\right)=\kappa_{0}\exp\left(-|z|/h\right)$,
where the absorption coefficient $\kappa_{0}$ (kpc$^{-1}$) at the
galactic plane and the vertical scale height $h$ are varied.

\begin{figure}[!t]
\begin{centering}
\includegraphics[clip,scale=0.4]{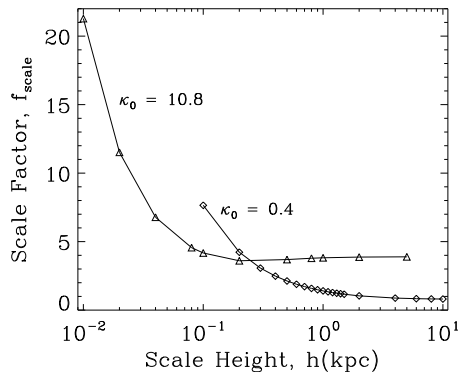}
\par\end{centering}

\caption{\label{fscale} Variation of the scale factor, $f_{{\rm scale}}\equiv\sum I_{{\rm data}}/\sum I_{{\rm model}}$,
for the exponential disk model with the absorption coefficient $\kappa_{0}=0.4$
kpc$^{-1}$ and $10.8$ kpc$^{-1}$.}

\end{figure}

We first confirmed that the Monte Carlo models with $h\ll1$ kpc are
consistent with the above thin slab models, and their total H$\alpha$
intensities increase linearly with the scale height. At an optical
thin limit of $\kappa_{0}\ll1$ kpc$^{-1}$, the morphologies of the
models are basically the same as the case of no absorption $(\kappa_{0}=0)$,
just as in thin slab models. Sample models are shown in Figure \ref{thick_models},
together with the intensity distributions of the models, observed
data in the second row, and correlation plots in the third row. The
scaling factors, defined by $f_{{\rm scale}}\equiv\sum I_{{\rm data}}/\sum I_{{\rm model}}$,
were multiplied by the models to match the overall H$\alpha$ brightness.
Unlike thin slab models, the total H$\alpha$ intensity is saturated
at $h\sim1/\kappa_{0}$ as the scale height increases since the amounts
of ionizing photons at high $|z|$ declines rapidly and the H$\alpha$
emission is dominated by the contributions from low $|z|\lesssim1/\kappa_{0}$.
Therefore, the scaling factor $f_{{\rm scale}}$ scaturates as $h$
increases. The trend is shown in Figure \ref{fscale} for the models
with $\kappa_{0}=0.4$, and 10.8 kpc$^{-1}$. As $\kappa_{0}$ increases,
the H$\alpha$ morphology becomes strongly concentrated near the \ion{H}{2}
regions, and the intensity distribution becomes wider as its low-intensity
side is elongated. The elongation of the low-intensity side of the
intensity histogram is shown in the blue curve of Figure \ref{thick_models}(d),
in which no $f_{{\rm scale}}$ was multiplied. For a given absorption
coefficient, the mophological coincidence of the thick disk models
with the observation is better than that of the thin slab models,
as the H$\alpha$ emission from the thick disk models are from more
extended regions. In other words, the intensity distribution is narrower
than that for the thin slab models. This can be easily noticed by
comparing the model of $\kappa_{0}=1$ kpc$^{-1}$ in Figure \ref{model_samples}
and that of $\kappa_{0}=2$ kpc$^{-1}$ in Figure \ref{thick_models}.
Even with a higher absorption cofficient, the thick disk model is
less concentrated near the \ion{H}{2} regions than the thin slab
model. As for thin slab models, we fitted the correlation between
the observed and model intensities with a power-law and plotted the
best fit curves in red lines and one-to-one curves in blue lines in
the third row of Figure \ref{thick_models}. We found that the models
with $\kappa_{0}=0.4-0.8$ kpc$^{-1}$ predict the observed data nicely,
and the scale height required to explain total H$\alpha$ intensity
is $\sim1-2$ kpc$^{-1}$.

The problem of the extremely low absorption rate remains unsolved
with the exponential disk models, although four times higher $\kappa_{0}$
are obtained. Assuming the same \ion{H}{1} column density over 0.1
kpc, as for the thin slab models, we obtained the effective absorption
cross-section $\sigma_{{\rm eff}}=(0.9-1.9)\times10^{-22}$ cm$^{2}$,
as low as $\sim10^{-5}$ of the photoionization cross-section. The
obtained scale height is consistent with that from thin slab models.
We note again that the scale height would be $\sim1-4$ kpc, considering
that the present results may be overestimated by up to a factor of
$2$.

\section{Opacity of the turbulent ISM to the Lyc}

We now examine whether such a small attenuation coefficient obtained
in \S3 is in accord with the current knowledge on the ISM density
structure. What we measure from observation is the average column
density $\left\langle N\right\rangle $ or optical depth $\left\langle \tau\right\rangle $
because of the finite spatial resolution. In the turbulent medium
showing the lognomal nature of column density, the effective optical
depth that the radiation field suffers, given an average optical depth
$\left\langle \tau\right\rangle $, can be calculated by \[
\tau_{{\rm eff}}=-\ln\int e^{-\tau}P(\ln\tau)d\ln\tau,\]
where the lognormal distribution is \[
P\left(\ln\tau\right)=\frac{1}{\sqrt{2\pi}\sigma_{\ln\tau}}\exp\left[-\frac{(\ln\tau-\ln\tau_{0})^{2}}{2\sigma_{\ln\tau}^{2}}\right]\]
\citep{Fischera03}. Here, $\left\langle \ln\tau\right\rangle \equiv\ln\tau_{0}$
and $\sigma_{\ln\tau}$ are the mean value and the standard deviation
in logarithmic space. The mean values in linear and logarithmic scales
are related by $\ln\tau_{0}=\ln\left\langle \tau\right\rangle -\sigma_{\ln\tau}^{2}/2$,
where $\left\langle \tau\right\rangle $ is the mean in the linear
scale. The relative standard deviation (contrast) in the linear scale,
$\sigma_{\tau}/\left\langle \tau\right\rangle =\sigma_{\tau/\left\langle \tau\right\rangle }$,
is related to the standard deviation in the logarithmic scale by $\sigma_{\tau/\left\langle \tau\right\rangle }^{2}=\exp(\sigma_{\ln\tau}^{2})-1$.
The relative deviation of optical depth $\sigma_{\tau/\left\langle \tau\right\rangle }$
is always smaller than the density contrast $\sigma_{\rho/\left\langle \rho\right\rangle }$,
as the accumulation of the density along a path length would average
out the density contrast \citep{Fischera04}. Given the observational
or theoretical value of the column density contrast $\sigma_{\tau/\left\langle \tau\right\rangle }$,
we can, then, obtain the lognormal distribution that describes the
column density, and estimate the effective optical depth, $\tau_{{\rm eff}}$,
of the Lyc using these relationships.

\begin{figure}[!t]
\begin{centering}
\includegraphics[clip,scale=0.4]{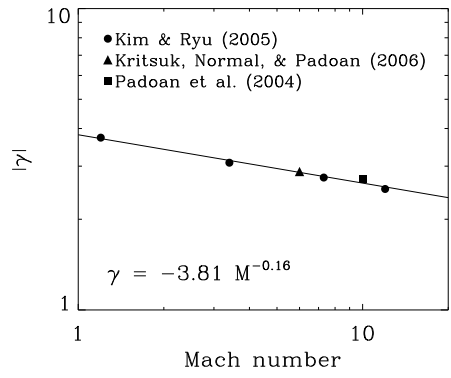}
\par\end{centering}

\caption{\label{mach_slope} Power-law slope versus the Mach number from numerical
simulations of the turbulent ISM, and the best-fit curve.}

\end{figure}

\begin{figure}[t]
\begin{centering}
\includegraphics[clip,scale=0.4]{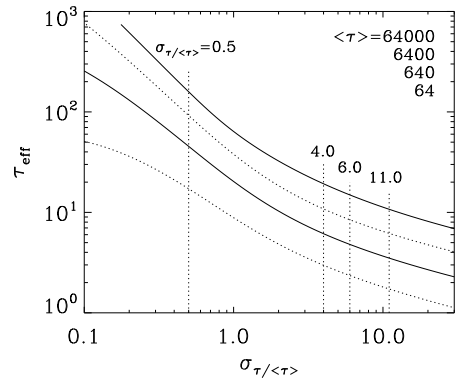}
\par\end{centering}

\caption{\label{tau_lognormal} Effective optical depth versus the column density
contrast for various average optical depths.}

\end{figure}

Observationally, the turbulent ISM has been usually investigated by
a power-law fit to the power spectra of the observable quantities
proportional to column densities. The power-law indexes have been
observed to be $\gamma\approx-2.4\sim-3.7$ from the warm and cold
ISMs in our Galaxy and the Small Magellanic Cloud \citep{Armstrong81,Crovisier83,Schlegel98,Stanimirovic00,Miville-Deschenes07}.
The power-law indexes can be related to the density contrast $\sigma_{\rho/\left\langle \rho\right\rangle }$
by combining two relationships between the Mach number $M$ and $\gamma$,
and between $M$ and $\sigma_{\rho/\left\langle \rho\right\rangle }$.
We combine the numerical results of \citet{Kim05}, \citet{Kritsuk06},
and \citet{Padoan04}, and find a relation between the Mach number
and the power-law index of the density power spectrum,\[
\gamma=(-3.81\pm0.07)M^{(-0.16\pm0.01)},\]
as shown in Figure \ref{mach_slope}. From this, we estimate $M\approx1-12$
for the observed $\gamma$. \citet{Fischera03} found that the empirical
reddening law for starburst galaxies is consistent with $M\approx1.3-22$.
\citet{Padoan97} found that $\sigma_{\rho/\left\langle \rho\right\rangle }\approx0.5M$
through numerical simulations. We then obtain the upper limit on the
column density contrast of $\sigma_{\tau/\left\langle \tau\right\rangle }\le\sigma_{\rho/\left\langle \rho\right\rangle }\approx0.5-11$.
In Figure \ref{tau_lognormal}, we show $\tau_{{\rm eff}}$ as a function
of $\sigma_{\tau/\left\langle \tau\right\rangle }$ for various $\left\langle \tau\right\rangle $.

Although, in real situations, the \ion{H}{1} column density is an
order of $\sim10^{20}-10^{21}$ cm$^{-2}$ over 0.1 kpc scale, we
calculate the effective optical depths for various column densities
to demonstrate how small an absorption coefficient is required by
the ``standard'' photoionization scenario. Assuming \ion{H}{1} column
densities of $N_{{\rm HI}}=10^{20}$, $10^{19}$, and $10^{18}$ cm$^{-2}$
over $0.1$ kpc, we have $\left\langle \tau\right\rangle =N_{{\rm HI}}\sigma_{{\rm photo}}$
(1 kpc/0.1 kpc) $\sim$ 64,000, 6400, and 640 over a distance of 1
kpc, respectively. We then obtain effective absorption coefficients
of $\kappa_{{\rm eff}}\gtrsim11-159$, $6.2-92$, and $3.5-45$ kpc$^{-1}$
from Figure \ref{tau_lognormal}, for $\sigma_{\tau/\left\langle \tau\right\rangle }\approx0.5-11$.
Note that these values are the lower limits for various \ion{H}{1}
column densities, but are still much larger, by several orders of
magnitude, than the $\kappa_{0}\approx0.4-0.8$ kpc$^{-1}$ required
to account for the DIG by the ``standard'' photoionization model.

Recently, \citet{Miville-Deschenes07a} found that $\sigma_{\rho/\left\langle \rho\right\rangle }\sim0.8$
for the warm neutral medium based on the kinematic \ion{H}{1} observation.
\citet{Hill08} showed for the first time that the H$\alpha$ intensity
PDF of our Galaxy is well represented by a lognormal distribution
and found that $M=1.3-2.4$ through comparison with the turbulence
simulations. We performed the same fit, as shown in Figure \ref{m51_map}(b)
and obtained a width of the lognormal distribution, $\sigma_{\ln I}\sim0.36\pm0.03$
dex, $\sim1.5$ times wider than that of our Galaxy. We obtain $M\lesssim6.5-7.3$
for the DIG of M 51 by comparing the result with Table 1 in \citet{Hill08}.
Assuming $M=8$ and $\sigma_{\tau/\left\langle \tau\right\rangle }\approx4$,
we have much stronger constraints on the absorption coefficient of
$\kappa_{{\rm eff}}\gtrsim$ 19.3, 10.8, and 6.1 kpc$^{-1}$ for $N_{{\rm HI}}=10^{20}$,
$10^{19}$, and $10^{18}$ cm$^{-2}$ over 0.1 kpc, respectively.

The models calculated with these absorption coefficients yield H$\alpha$
morphologies significantly different from the observed one. This implies
that the Lyc photons leaked from the \ion{H}{2} regions are immediately
absorbed by the diffuse ISM, even if it has escaped out of the \ion{H}{2}
regions. For comparison, we calculated the Mach numbers required to
have the effective absorption coefficient $\kappa=0.8$ kpc$^{-1}$,
assuming that the same relation between the standard deviation of
column density and the Mach numer is still valid in this extreme case:
$M>2.0\times10^{5}$, $1.8\times10^{4}$, and $1.6\times10^{3}$ for
the media having $N_{{\rm HI}}=10^{20}$, $10^{19}$, and $10^{18}$
cm$^{-2}$ over $0.1$ kpc, respectively.

\section{Discussion}

\subsection{Vertical Profile}

\begin{figure}[t]
\begin{centering}
\includegraphics[clip,scale=0.5]{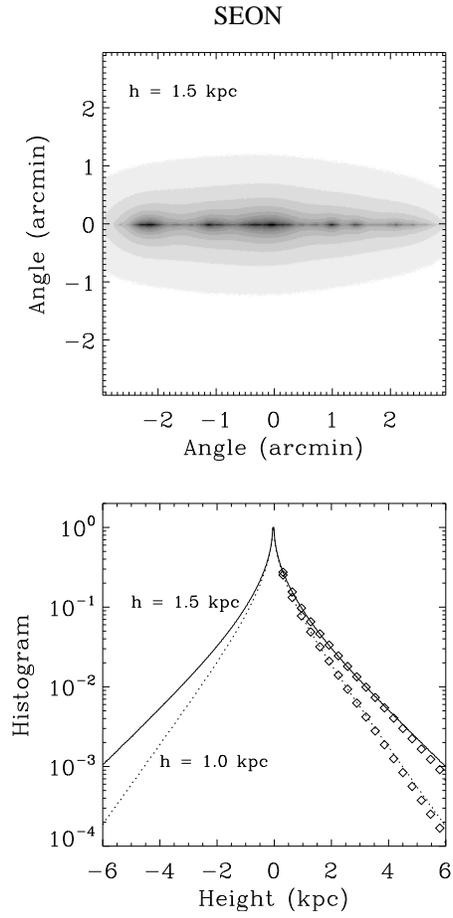}
\par\end{centering}

\caption{\label{edge-on}(a) Model for the galaxy M 51 when it is viewed edge-on,
assuming $\kappa_{0}=0.4$ kpc$^{-1}$ and $h=1.5$ kpc, (b) the vertical
profile of the H$\alpha$ intensity for two models: $\kappa_{0}=0.4$
kpc$^{-1}$, $h=1.5$ kpc, $\kappa_{0}=0.4$ kpc$^{-1}$, and $h=1.0$
kpc. Here, the diamond symbols denote two-exponential fits for $|z|>0.3$
kpc.}

\end{figure}

By comparing the H$\alpha$ emission data with models, we found that
the vertical scale height of the face-on galaxy M 51 must be $\sim1-2$
kpc. Even though this value is obtained from the problematic absorption
coefficient, we note that the scale height is consistent with the
results not only from our Galaxy, but also from external edge-on galaxies
\citep{Reynolds89,Rand1996,Collins2001}. It might also be noteworthy
to examine how the galaxy M 51 may look when the galaxy is viewed
edge-on. In Figure \ref{edge-on}, we show the H$\alpha$ map, assuming
$\kappa_{0}=0.4$ kpc$^{-1}$ and $h=1.5$ kpc, which are expected
when the galaxy is viewed edge-on, and the vertical profile of H$\alpha$
image. The vertical profile of the model with the $\kappa_{0}=0.4$
kpc$^{-1}$ and $h=1.0$ kpc are shown as well for comparision. The
H$\alpha$ intensity from \ion{H}{2} regions are not shown in the
figure.

It is surprising that the simulated vertical H$\alpha$ profile can
be represented by a function consisting of two or more-exponential
components. Observationally, the profile might be recognized by two-exponential
components because of low surface brightness at high $|z|$. If the
Lyc photons travel only perpendicularly to the galactic disk, the
H$\alpha$ profile would have the same scale-height as that of the
exponential disk. However, the H$\alpha$ radiation observed at a
point with vertical height $z$ consists of the contributions from
the Lyc photons that have traveled longer distances, i.e., $(r^{2}+z^{2})^{1/2}>|z|$,
where $r$ is the projected distance onto the galactic plane from
the Lyc source. Therefore, the H$\alpha$ intensity declines faster
than that naively expected from the assumed vertical height of the
exponential disk. The smaller the vertical distance, the more important
the effect becomes. Beyond the height correponding to the system size
$\sim10$ kpc, the vertical profile converges to that of the disk,
as the distances traveled by the Lyc become close to the vertical
height, i.e., $(r^{2}+z^{2})^{1/2}\approx|z|$. The trend is noticeable
in Figure \ref{edge-on}(b). We then have a vertical profile that
resembles two-exponential components, within a few kpc. The profiles
in Figure \ref{edge-on}(b) are fitted with $0.19e^{-z/1.1}+0.33e^{-z/0.32}$
for $h=1.5$ kpc, and $0.23e^{-z/0.80}+0.32e^{-z/0.26}$ for $h=1.0$
kpc, and the fit profiles are denoted by diamonds. We have masked
the regions at $|z|<0.3$ kpc in fitting the vertical profile to minimize
the effect due to the assumption that the \ion{H}{2} regions are
located in the galactic plane ($z=0$). The vertical height of the
smaller exponential component is expected to depend not only on the
vertical profile of \ion{H}{1} gas, but also of the details of the
vertical extents of the \ion{H}{2} regions.

The H$\alpha$ vertical emission profiles of edge-on galaxies are
known to be well represented by two exponential components \citep{Rand1997,Miller2003}.
\citet{Wang1997} identified the two components kinematically through
the observation of face-on galaxies, including M 51, a ``quiescent''
DIG with a high ionization state and small scale height (few hendred
pc), and a ``disturbed'' DIG with a high ionization state and moderate
scale height (0.5--1 kpc). The origin and evolution of the extended,
or extraplanar, DIG seen in edge-on galaxies are not clear, but two
alternatives are generally considered \citep[e.g., ][]{Heald2006,Heald2007}.
The gas is generally considered to participate in a star formation-driven
disk-halo flow, such as that described in the fountain or chminey
models \citep{Shapiro1976,Norman89}. Alternatively, the gas could
have been accreted from the intergalactic medium or from companion
galaxies \citep[e.g., ][]{vanderHulst}. The vertical structure seen
in our models seems to support the star formation-related origin of
the extraplanar DIG. However, without kinematical information, as
was used in \citet{Heald2006,Heald2007}, it is not easy to examine
the two-components predicted in these models with $\kappa=0.4$ kpc$^{-1}$
do correspond to the kinematically identified components, and/or to
determine how much of the DIG is associated with the quiescent component
and how much is contained in extended disturbed layers. In fact, it
is not only surprizing, but also very puzzling, that the models with
a problematic absorption rate predicts not only the apparent morpologies
of M 51, but also the two exponential components seen in edge-on galaxies.

\subsection{Density Stucture of the ISM}

\begin{figure*}[t]
\begin{centering}
\includegraphics[clip,scale=0.8]{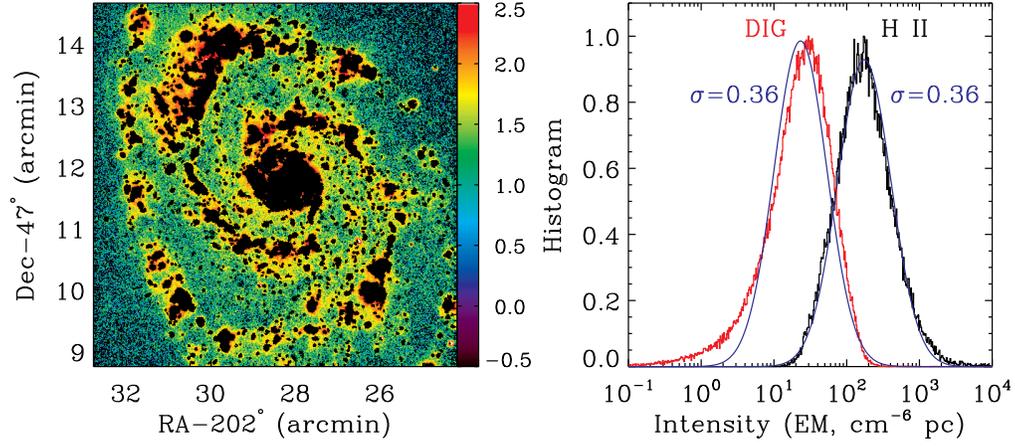}
\par\end{centering}

\caption{\label{m51_map}(a) H$\alpha$ image of M 51 in logarithmic scale
(EM, cm$^{-6}$ pc). The \ion{H}{2} regions were masked. (b) Histograms
of H$\alpha$ intensities for the DIG and \ion{H}{2} regions. Blue
curves denote the best-fit lognormal functions, as discussed in \S4
and \S5.}

\end{figure*}

\begin{figure*}[!t]
\begin{centering}
\includegraphics[clip,scale=0.49]{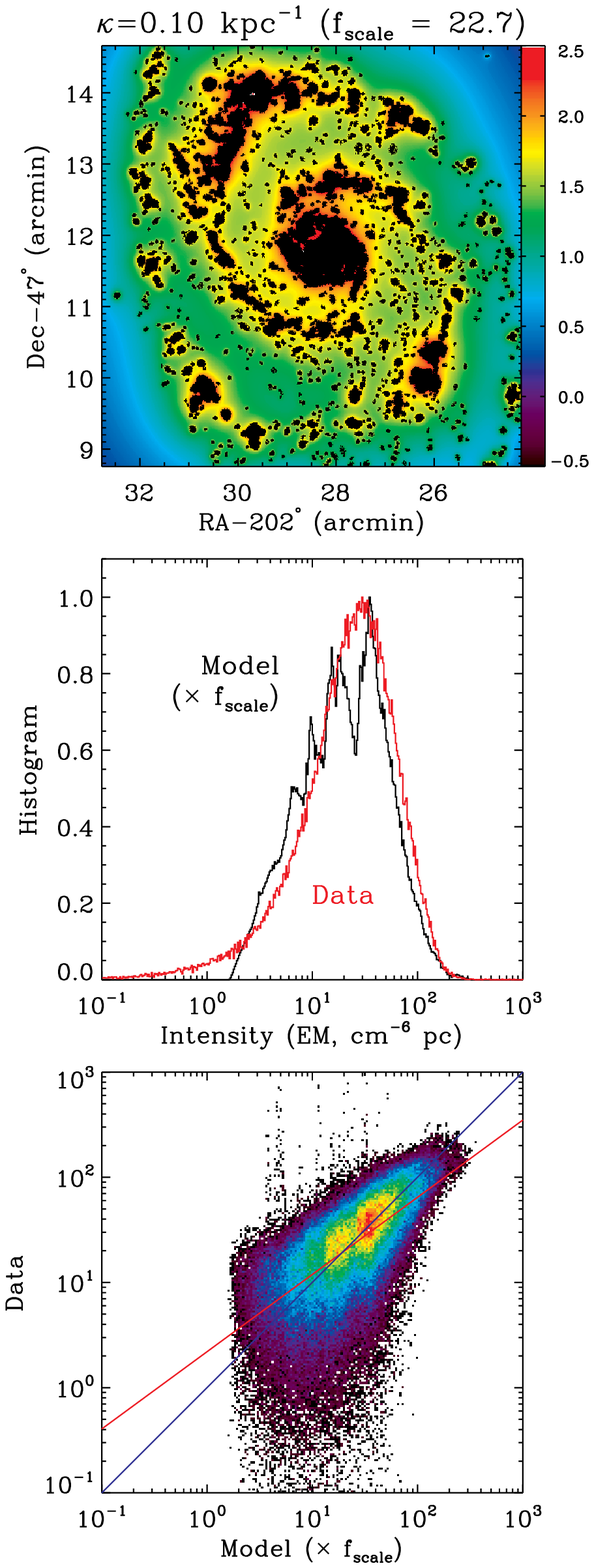}\includegraphics[clip,scale=0.49]{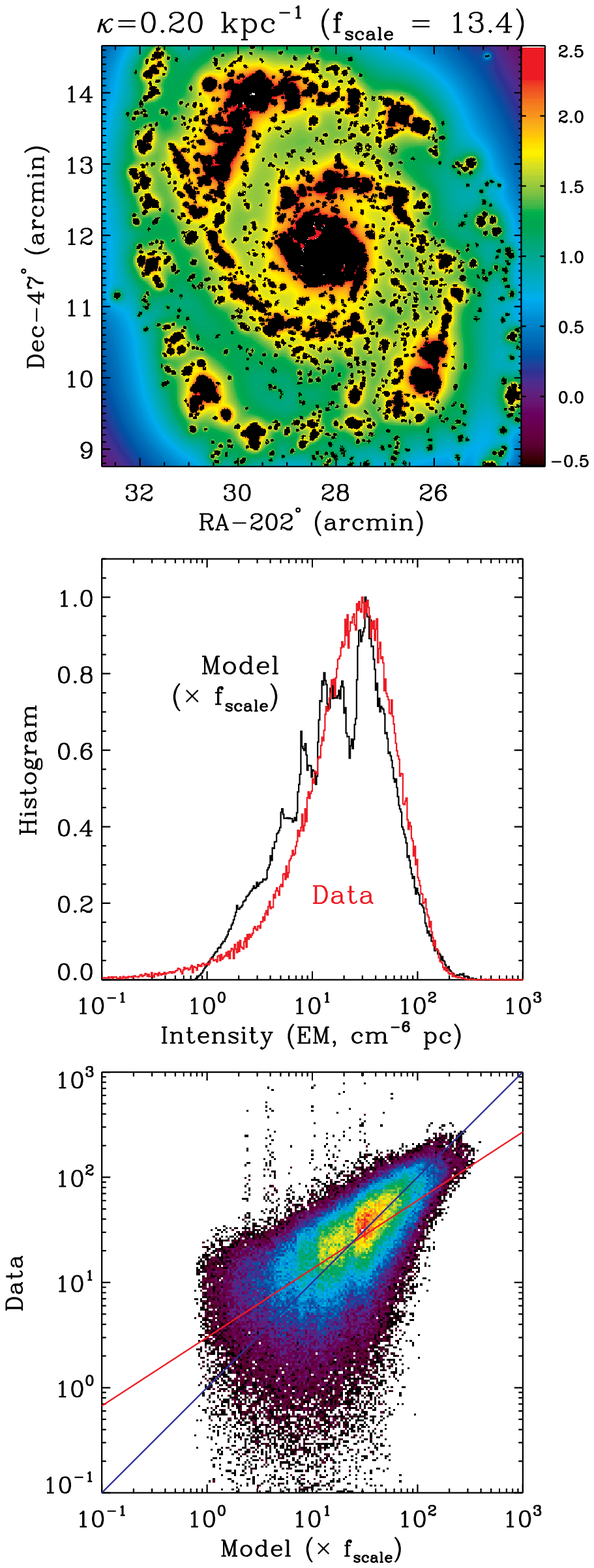}\includegraphics[clip,scale=0.49]{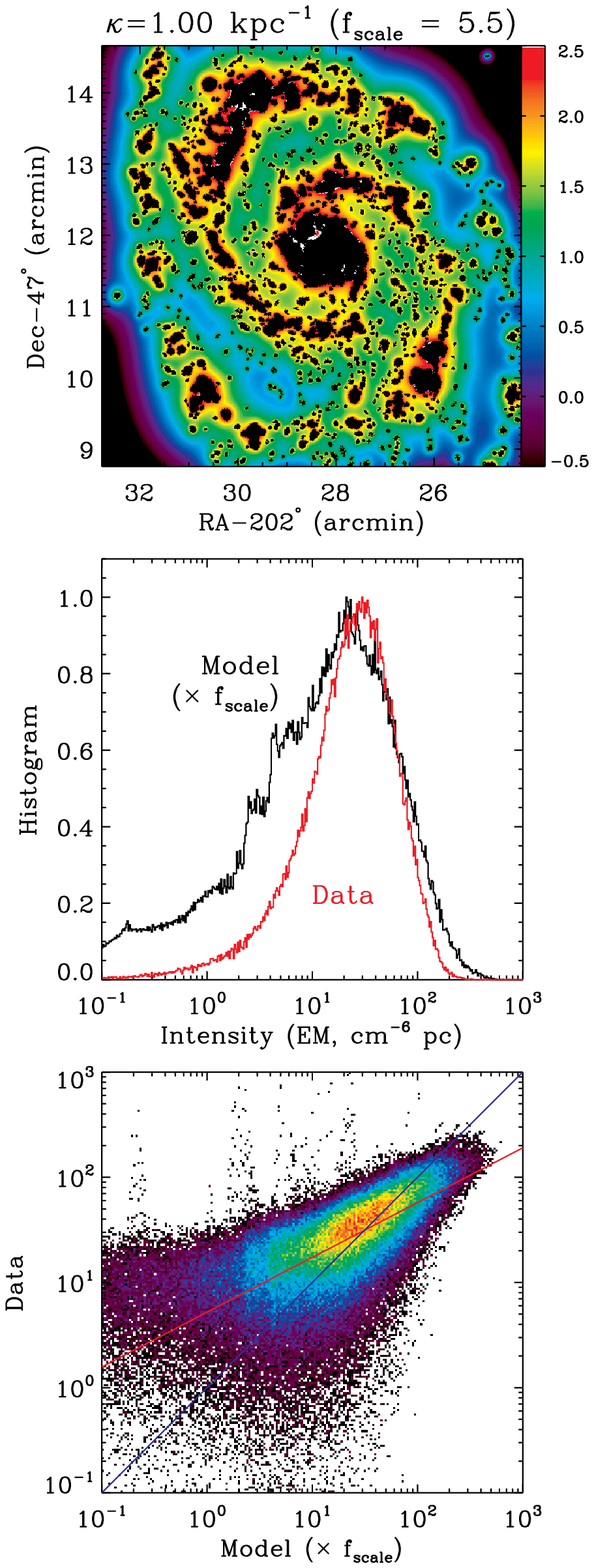}\includegraphics[clip,scale=0.49]{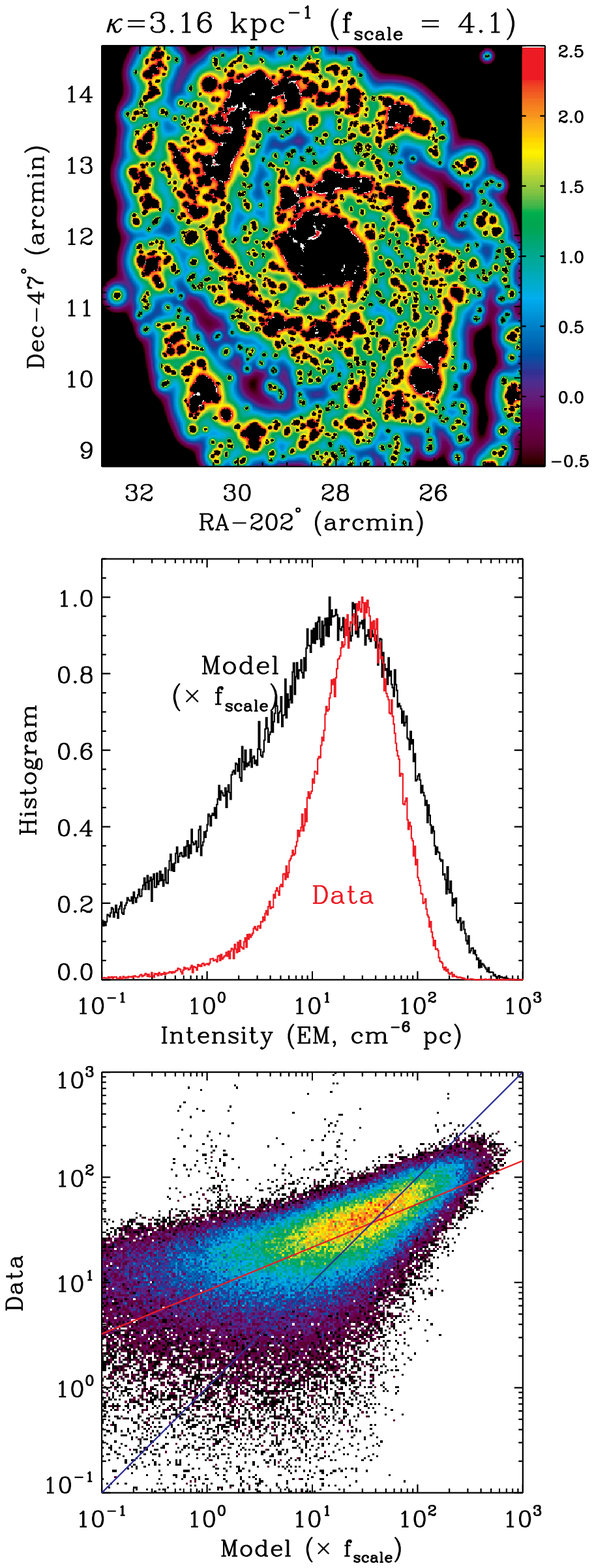}
\par\end{centering}

\caption{\label{model_samples}Thin slab models calculated for various absorption
coefficients. Predicted model images with the same color scales as
in Fig. \ref{m51_map}, intensity histograms of the model and observed
data, and correlation plots between the model and observed data are
shown in the first, second, and third rows, respectively.}

\end{figure*}

\begin{figure*}[!t]
\begin{centering}
\includegraphics[clip,scale=0.49]{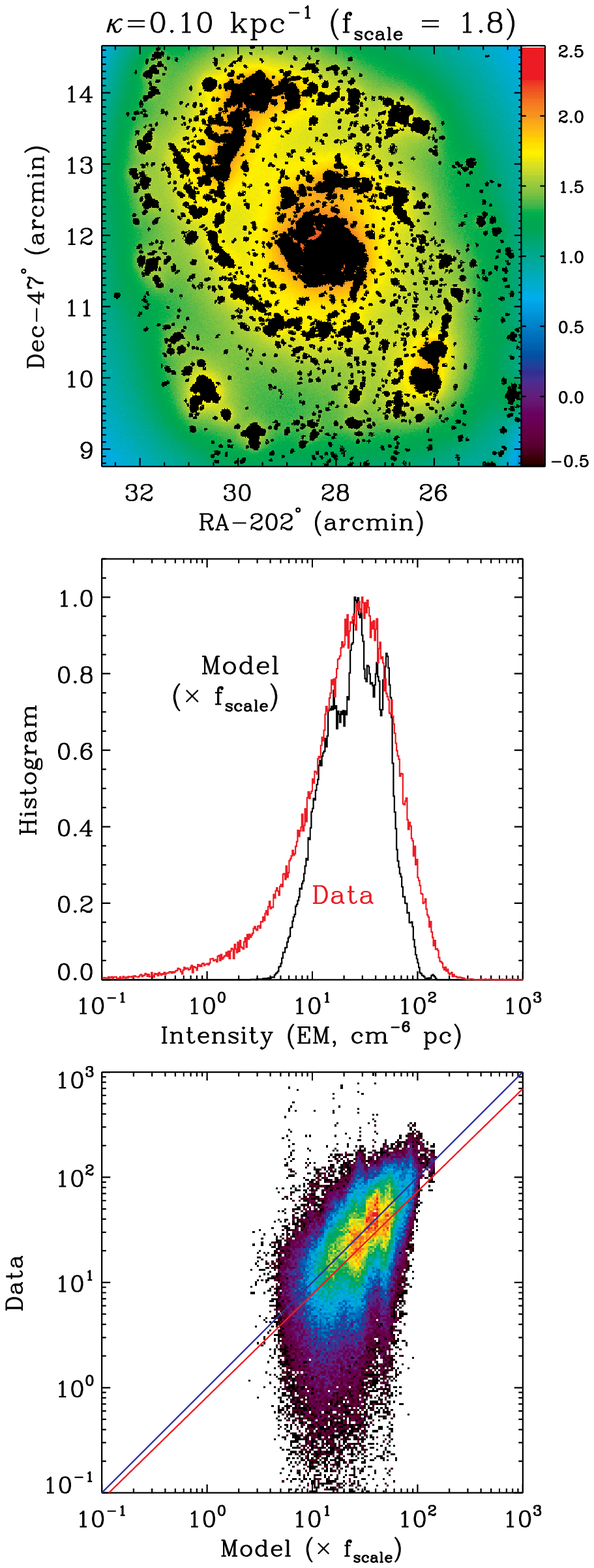}\includegraphics[clip,scale=0.49]{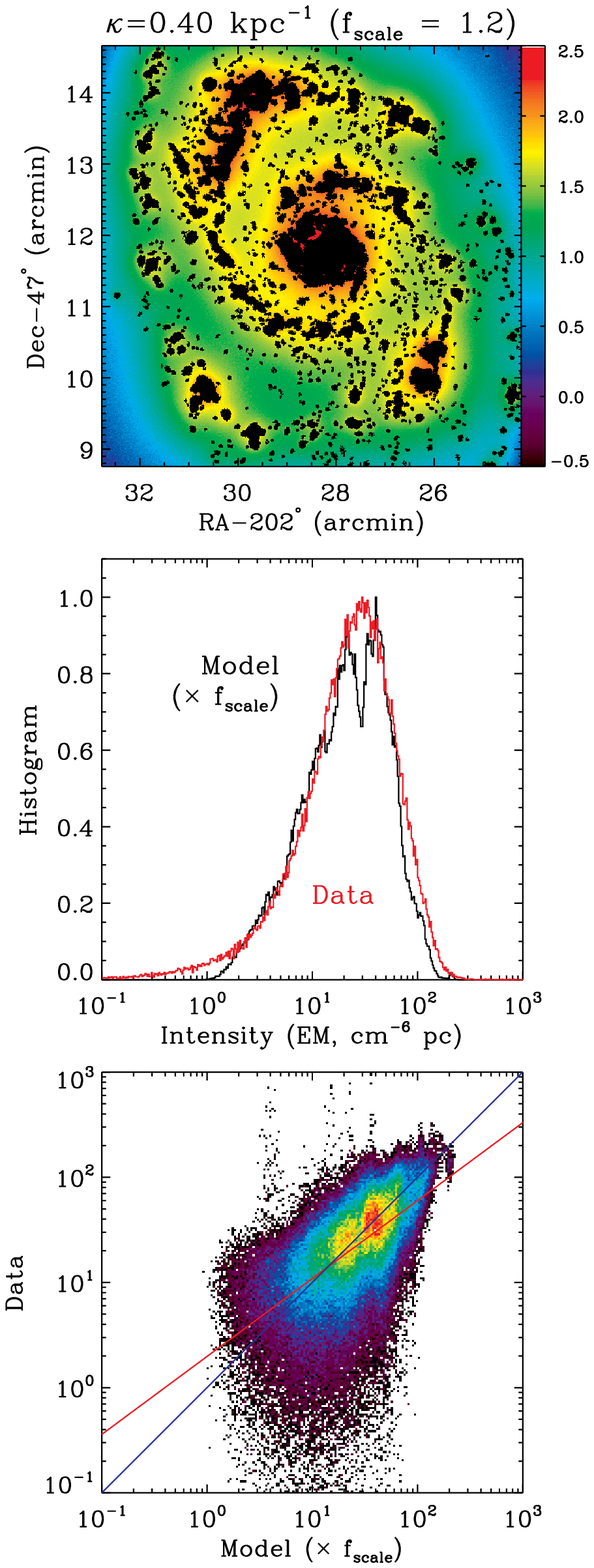}\includegraphics[clip,scale=0.49]{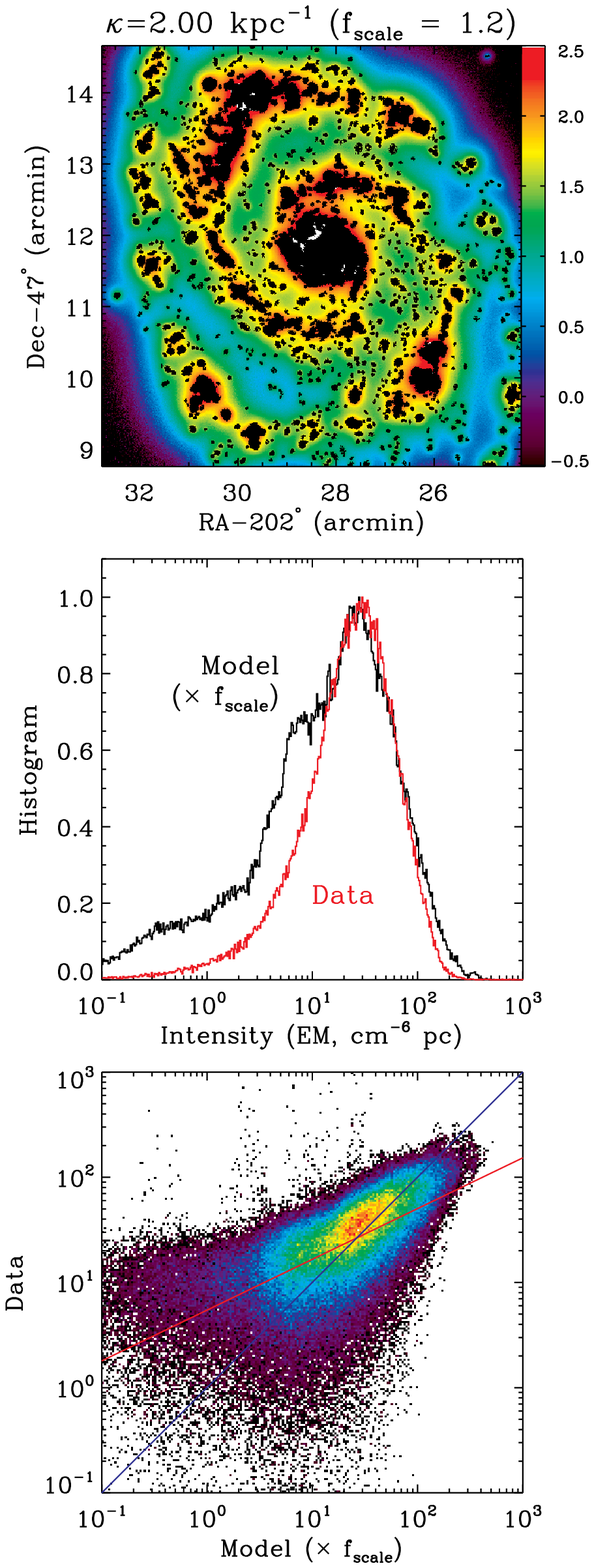}\includegraphics[clip,scale=0.49]{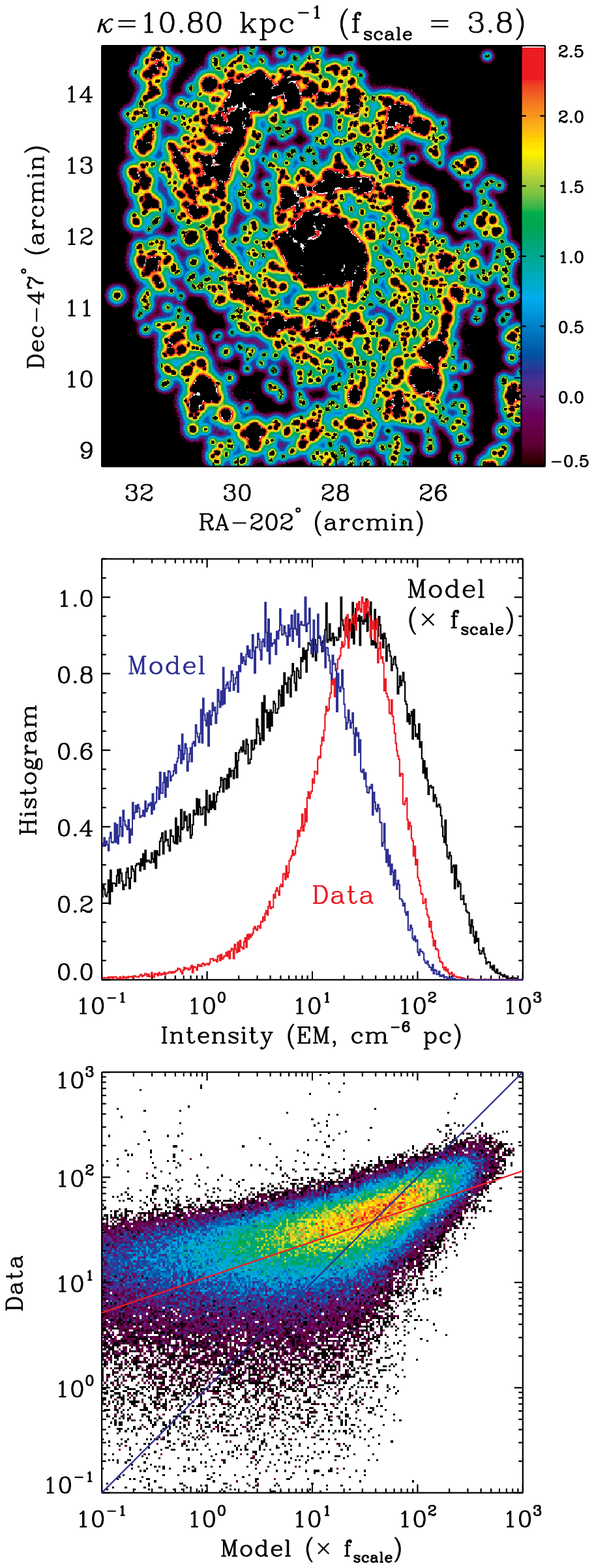}
\par\end{centering}

\caption{\label{thick_models}Exponential disk models calculated for various
absorption coefficients and scale height: (a) $\kappa_{0}=0.1$ kpc$^{-1}$
and $h=3$ kpc, (b) $\kappa_{0}=0.4$ kpc$^{-1}$ and $h=1.5$ kpc,
(c) $\kappa_{0}=2.0$ kpc$^{-1}$ and $h=1$ kpc, and (d) $\kappa_{0}=10.8$
kpc$^{-1}$ and $h=1$ kpc. Predicted model images with the same color
scales as in Fig. \ref{m51_map}, intensity histograms of the model
and observed data, and correlation plots between the model and observed
data are shown in the first, second, and third rows, respectively.}

\end{figure*}

As suggested by numerical simulations \citep[e.g., ][]{Vazquez94,Kim05},
the density distribution of turbulently disturbed ISM is represented
by a lognormal distribution, of which the width depends on how the
medium is turbulent. \citet{Padoan97} found that the observational
result of \citet{Lada1994}, which was that the variation of the stellar
extinction in dark clouds correlates with the mean extinction, is
consistent with the lognormal distribution of the dust density. \citet{Wada2000}
showed that the luminosity function of the \ion{H}{1} column density
in the Large Magellanic Cloud is lognormal. More recently, \citet{Hill08}
found that the H$\alpha$ emission measure perpendicular to the Galactic
plane from the DIG in the Milky Way is well fitted by a lognormal
distribution, with a standard deviation of $\sigma_{\ln I}\sim0.2$
dex. \citet{Berkhuijsen2008} showed that the PDFs of the average
density, not only of the DIG, but also of the \ion{H}{1} gas in our
Galaxy, are close to lognormal, with a standard deviation $\sigma_{\left\langle n\right\rangle }\sim0.3$
dex. In Figure \ref{m51_map}(b), we showed that the PDFs of the H$\alpha$
intensity, i.e., emission measures, from the DIG and \ion{H}{2} regions
in M 51 are well-represented by a lognormal function. The standard
deviation obtained from the DIG in M 51 is a bit higher than those
of our Galaxy.

\citet{Wang1997} studied kniematics of the DIG in M 51, together
with other galaxies and found that the DIG is more disturbed kinematically
than the gas in the giant \ion{H}{2} regions. We, thus, compare the
widths of H$\alpha$ intensity distributions of the DIG and \ion{H}{2}
regions in Figure \ref{m51_map}(b) and found no significant difference
between them, except an extended tail toward low intensities shown
in the DIG only. However, it is unclear whether the extended tail
might be related to the more disturbted component seen by \citet{Wang1997}.

In \S4, we found that an unrealistically small absorption, even smaller
than dust-extinction, is necessary to explain the global features
of the DIG with the Lyc leakage from the \ion{H}{2} regions. The
turbulent ISM with lognormal density PDF can be much more transparent
to the Lyc, up to an order of $1\sim3$ depending on the width of
the PDF, compared to homogeneous ISM, as shown in Figure \ref{tau_lognormal}.
However, such a small attenuation found here cannot be reconciled
with these values, as discussed in \S4. In addition, the attenuation
rate is not consistent even with the dust optical depth of $\tau_{{\rm dust}}\sim13$
at 912\AA\ ($A_{V}\sim3$) obtained by comparing H$\alpha$ and Pa$\alpha$
emissions measured toward the \ion{H}{2} regions of M 51 \citep{Quillen01,Scoville01}.
Such large amounts of dust surrounding \ion{H}{2} regions should
absorb Lyc photons immediately in the vicinities of the \ion{H}{2}
regions.

Radiative transfer models have been successful in explaining the dust
scattering and absorption for our Galaxy and external galaxies with
reasonable amounts of dust, consistent with the observed amounts of
interstellar dust \citep[e.g., ][]{Schiminovich2001,Bianchi2007}.
Unless the topology of the interstellar dust and neutral hydrogen
are completely different and decoupled from each other, the radiative
transfer of the Lyc should not differ significantly from that expected
from observed amounts of neutral hydrogen. In their study of dust
scattering in clumpy media, \citet{Witt1996} showed that the effective
optical depth of a system is mainly determined by the covering factor,
i.e., the fraction of $4\pi$ steradian of a solid angle filled by
clumps as seen by the source rather than by the individual optical
depth of clumps. The transmission of the Lyc is also mainly determined
by the covering factor of dilute regions seen by the source. Such
a large covering factor of low density regions may be achieved by
fountain flow or chimneys around OB associations \citep{Shapiro1976,Norman89}.
Filaments seen in edge-on galaxies may originate from the star formation-driven
structures. Meanwhile, \citet{Rossa2004} presented high spatial resolution
(5 pc) observations of the extraplanar DIG in the edge-on galaxy NGC
891 and found no clear broadening of the filamentary structures toward
high $|z|$ expected from individual chimneys. They, therefore, concluded
that the chimney scenario is most likely not responsible for the gas
and energy transport into the halo. In addition, \citet{Wang1997}
found that M 51 has a pronounced ``diffuse'' H$\alpha$ emission based
on its high value of mean/rms of the H$\alpha$ surface brightness.
\citet{Dove00} showed that the dense swept up shell of material by
the dynamical action of supernovae and stellar winds would trap ionization
photons within superbubbles. Given the diffuseness or smoothness of
the H$\alpha$ emission and the approximate equality of H$\alpha$
luminosities of the DIG and \ion{H}{2} regions, the ISM should have
an extremely unrealistic configuration, completely evacuated straight
thin holes extending over the order of 1 kpc, randomly covering $\sim1/2$
of the $4\pi$ solid angle seen by the sources.

The atomic ISM may be divided into two phases, except the hot gas
of temperature $\sim10^{6}$ K: the neutral hydrogen gas (or clouds)
and the DIG (or interclouds) phases. Our aim was to investigate the
situation in which the stellar Lyc radiation propagates into the ``initially''
neutral ISM, ionize some portion of the gas, and maintain two phases.
Since the ISM is not fully ionized, the \ion{H}{1} gas prevents the
propagation of the Lyc photons to maintain the ionization state of
the DIG. However, it would be worthwhile to imagine an ISM in which
the \ion{H}{1} gas phase has no effect on the propagation of the
stellar Lyc photons and the freely transmitted Lyc radiation maintains
the ionization state of the DIG. In other words, assume that the \ion{H}{1}
gas phase is completely evaculated somehow, and the DIG is left alone.
Then, the radiative tranfer in the medium will be determined by the
observed property of the DIG that is deduced from the H$\alpha$ intensity
distribution in \S4. Adopting $M=8$, $\sigma_{\tau/\left\langle \tau\right\rangle }=4$
from the DIG of M 51, and $N_{{\rm HI}}\sim1\times10^{20}$ cm$^{2}$
over $\sim1$ kpc estimated in the Galactic plane \citep{Reynolds91},
the effective absorption coefficient is $\sim10.8$ kpc$^{-1}$. The
model for these parameters is shown in Figure \ref{thick_models}(d).
The model shows a highly concentrated H$\alpha$ profile near the
\ion{H}{2} regions and predicts only about $1/f_{{\rm scale}}\sim26$\%
of the observed H$\alpha$ luminosity. Since the relative deviation
of the column density is an upper limit, the amount of H$\alpha$
that can be produced in this medium is even smaller, i.e., $\lesssim26$\%.
A more stringent constraint would be obtained, assuming a Mach number
of $M=\sim1.2-2.4$ estimated from our Galaxy \citep{Hill08}. Therefore,
the OB associations in the \ion{H}{2} regions are not likely to be
a main source in ionizing the DIG. The only way to resolve this problem
would be to assume alternative ionizing sources outside of \ion{H}{2}
regions, in-situ of the DIG. If this is the case, the similarity in
the H$\alpha$ morphologies between the models and the observational
data may be a coincidence.

\section{Concluding Remarks}

We investigated whether the global features of the DIG can be explained
in the context of the ``standard'' photoionization model. The most
important result of this study is that we quantified the topology
or ``porosity'' of the ISM required by the ``standard'' photoionization
scenario in the contexts of the lognormal PDF. We found that even
without the \ion{H}{1} gas phase, the DIG phase cannot be maintained
by the stellar Lyc radiation alone, implying that main ionizing source
is not the OB associations in the \ion{H}{2} regions. Since the Lyc
that leaked out of the \ion{H}{2} regions can account for, at most,
only $\lesssim26$\% of the H$\alpha$ luminosity of the DIG, the
alternative sources should play a significant role in ionizing the
DIG, especially in the interarm regions. Even with the mophological
coincidence of the Lyc leakage models with the H$\alpha$ data, we
cannot rule out the possibility that alternative ionizing sources
are generally correlated with star formations and, thus, with the
\ion{H}{2} regions, as various gas and dust components are well correlated
with each other, and the sources are playing a significant role in
ionizing the DIG. The strong correlation between the DIG and \ion{H}{2}
regions does not necessarily indicate the direct association of the
DIG with the \ion{H}{2} regions.

Other authors have arrived at similar conclusions, both in terms of
producing the amount of ionized gas observed and in producing the
observed spectral signature of the DIG, that alternative sources other
than hot stars in the \ion{H}{2} regions are needed in heating and/or
ionizing the DIG. \citet{Rand1990} estimated that the total energy
input by chimneys into the halo to account for the DIG of NGC 891,
a galaxy that in many ways is very similar to our own, is a factor
10--20 higher than the estimates for our Galaxy. Although the supernova
rate is likely to be a few times higher than in our Galaxy, the required
superbubble formation rate was inconsistent with the observations.

The emission-line ratios of the DIG are significantly different from
those of typical \ion{H}{2} regions. In the DIG, {[}\ion{Si}{2}{]}
$\lambda6716$/H$\alpha$ and {[}\ion{N}{2}{]} $\lambda6583$/H$\alpha$
are enhanced relative to the \ion{H}{2} regions \citep{Haffner1999,Ferguson1996b,Rand1997,Greenawalt,Collins2001,Hoopes2003,Voges2006}.
The pure photoionization models in diluted media predict the enhancement
very well. However, the models predict that the ratio {[}\ion{Si}{2}{]}/{[}\ion{N}{2}{]}
should increase with $z$, in contrast to the observations that the
ratio versus the distance from the \ion{H}{2} regions remains nearly
constant in the DIG \citep{Haffner1999,Collins2001}. The behaviors
of these low ionization lines can be explained by the gas temperature
increase alone, without having a secondary source of ionization \citep{Haffner1999}.
On the other hand, the ratio {[}\ion{O}{3}{]} $\lambda5007$/H$\alpha$
cannot be fully explained by a temperature increase alone \citep{Collins2001}.
According the ``standard'' photoionization scenario, the ratio {[}\ion{O}{3}{]}/H$\alpha$
should decreases with $z$ or the distance from the \ion{H}{2} regions
in a photoionization scenario. However, the opposite trend or the
relative constant with $z$ is seen not only in edge-on galaxies \citep{Rand1998,Collins2001},
but also in face-on galaxies including M 51 \citep{Wang1997,Hoopes2003,Voges2006}.
One possible way to explain this high line ratio is to assume a secondary
ionization source that contributes a small but increasing fraction
of the H$\alpha$ with $z$. \citet{Collins2001} explored the possibility
that ionization by shocks \citep{Shull1979} or turbulent mixing layers
\citep{Slavin1993} is responsible for some of the DIG emission of
the edge-on galaxies NGC 4302 and UGC 10288. They found that the composite
models, of which up to $\sim16-45$\% of the H$\alpha$ arises from
shocks, better reproduce the variations of line ratios than the photoionization
models alone, although no model reproduces the line ratios to within
the errors. Simliar conclusions were obtained for other edge-on galaxies
by \citet{Rand1998} and \citet{Miller2003b}. \citet{Hoopes2003}
tested the feasibility of ``leaky \ion{H}{2} region'' models, in
wich the transmitted ionizing continua of density-bounded \ion{H}{2}
regions power the DIG, and concluded that other ionization or heating
mechanisms play a role in ionizing the DIG. In addition, the \ion{He}{1}
$\lambda$5876/H$\alpha$ line intensity ratio in the DIG of our Galaxy
is found to be significantly less than that observed in \ion{H}{2}
regions, challenging the O star ionization models \citep{Reynolds1995}.

Another scenario investigated by \citet{Hoopes00}, \citet{Hoopes01}
is that field OB stars ionize some of the DIG. They measured the H$\alpha$/FUV
intensity ratios of ten spiral galaxies, and concluded that late OB
stars in the field are indeed important ionization sources. However,
\citet{Voges2006} concluded that field star ionization models do
not fit the data of M 33 well and that about 30\% leakage from \ion{H}{2}
regions is necessary. Accretion from the intragalactic medium or companion
galaxy would be an attractive alternative for the origin of, at least
some of, the DIG. \citet{Heald2006,Heald2007} compared the kinematics
of optical emission lines obtained from two edge-on galaxies NGC 891
and NGC 4302 with an ballistic model of a galactic fountain model
\citep{Collins2002} and found a strong disagreement between the observed
kinematics and those predicted by the model. As they noted, galaxies
may reside in halos populated by a various mixture of accreting gas
and star formation-driven gas. However, further analyses are required
to quantitatively estimate their contributions in ionizing the DIG.

\clearpage{}%

\end{document}